\documentstyle[aps,prb,epsf,floats]{revtex}
\begin{document}
\twocolumn[\hsize\textwidth\columnwidth\hsize\csname@twocolumnfalse%
\endcsname
\title{Low temperature relaxational dynamics of the Ising 
chain in a transverse field}
\author{Subir Sachdev}
\address{Department of Physics, Yale University\\
P.O. Box 208120,
New Haven, CT 06520-8120}
\author{A.P. Young}
\address{Department of Physics, University of California\\
Santa Cruz CA 95064}
\date{September 16, 1996}
\maketitle
\begin{abstract}
We present asymptotically exact results for 
the real time order parameter
correlations of  a class of $d=1$ Ising models in a transverse
field at low temperatures ($T$) on both sides of the
quantum critical point. The correlations are a
product of a $T$-independent factor 
determined by quantum effects, and a $T$-dependent relaxation function
which comes from a {\em classical} theory.
We confirm our predictions by a no-free-parameter comparison with
numerical studies on the 
nearest neighbor spin-1/2 model.
\end{abstract}
]

Real time, non-zero temperature ($T$), correlation functions of quantum many
body systems are directly related to the observables of many experiments.   For
strongly interacting systems, there are few quantitative results on
the relaxation and transport processes that are believed to occur at long times
at any $T\neq 0$. Monte Carlo and perturbative methods work best in imaginary
time, but the analytic continuation to real time is most dangerous, and often
fails, in the low frequency limit.  At
special conformally invariant points in dimension $d=1$, real 
time correlations describing
relaxation of an order parameter can be computed exactly.  Among systems in
arbitrary $d$, which are tuned across a quantum critical point by
a variable coupling, there is only one for which reliable results are
available: the  $d=1$ impenetrable Bose gas, whose correlators were determined
by a profound and sophisticated inverse scattering analysis~\cite{korepbook}.

In this paper, we study real time, $T\neq 0$ correlations of the $d=1$ Ising
model in a transverse field. We use a novel semiclassical method to obtain the
exact asymptotics of order 
parameter correlations in the two low $T$ regions on either side of
its quantum critical point.
Our main new result, given in Eqs.~(\ref{main_res_RC}), (\ref{main_res}) below,
is that in these low $T$ regimes,
the spin correlation function can be expressed as a product of two
factors. One, arising from quantum effects, gives the $T=0$ value of the
correlation function, and the other,
which comes from a {\em classical} theory, describes the effects of
temperature. 
Combined with earlier results~\cite{perk,ssising},
our new results
give a complete description of time-dependent
correlations in all the distinct limiting regions of the Ising model,
and exhibit, simply and clearly, the crossovers in the
roles of quantum and thermal fluctuations in the relaxational dynamics.

For the Ising chain with only nearest neighbor exchange, we also
use the free fermion
representation~\cite{lsm,mccoy}
to obtain accurate, numerical data~\cite{yr,snm} for real time spin
correlations for systems with up to 512 spins: these results are in
excellent agreement with the asymptotic theoretical results even at relatively
short times and distances.

We also use our semiclassical method to obtain
correlators of the $d=1$ impenetrable Bose gas in a certain low $T$ regime; our
results here are in agreement with earlier work~\cite{korepbook}, although our
approach is much simpler 
and more physically transparent. Our method should also apply
to other $d=1$ quantum models (like the non-linear sigma or
sine-Gordon) with an excitation gap.

\begin{figure}[htbp]
\epsfxsize=3in
\centerline{\epsffile{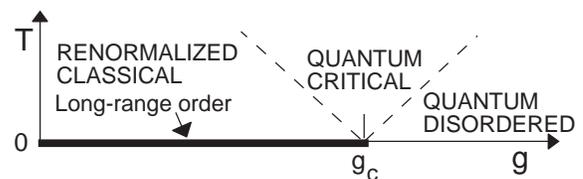}}
\caption{
Finite $T$ phase diagram of $H_I$. The dashed lines are crossover
boundaries at $T \sim |m c^2|$.
}
\label{fig1}
\end{figure}

We consider the Hamiltonian
\begin{equation}
H_I = -\sum_i \left( 
\sum_{\ell >0} J_{\ell} \sigma^z_i \sigma^z_{i+\ell} + g \sigma^x_i \right)
\label{ham}
\end{equation}
where $\sigma^z_i$, $\sigma^x_i$ are Pauli matrices on a chain of sites $i$,
$J_{\ell}$ $(> 0)$ are
short-ranged exchange constants, and $g$ $(>0)$
is the transverse field. The ground
state of $H_I$ is expected to have long range order with $N_0 \equiv \langle
\sigma^z \rangle \neq 0$ 
for all $g < g_c$, and a gap to all excitations for $g
\neq g_c$~\cite{lsm}, see Fig~\ref{fig1}.
Away from $g=g_c$, the low lying states consist of a stable
particle with energy $\epsilon_p$ at momentum $p$, and $n > 1$ particle
continua at higher energy. We also have $\epsilon_{-p} = \epsilon_p$, and
expect that $\epsilon_p$ has a minimum at $p=0$.
Consequently,
for small $p$ we
parameterize $\epsilon_p = ((m c^2)^2 + c^2 p^2 + {\cal O}(p^4))^{1/2}$, which
defines the `velocity' $c>0$ and the `mass' $m$; we choose $m >
0$ ($m < 0$) for $g < g_c$ ($g > g_c$). The critical point at $g=g_c$ is
described by a continuum theory with dynamic exponent $z=1$, and correlation
length exponent $\nu=1$;
hence $c \sim \mbox{constant}$ for $g$ near $g_c$, while
$m$ vanishes linearly, $|m| \sim |g-g_c|$, with possibly a different slope on
the two sides.

The model 
with $J_{\ell>1} = 0$ is integrable~\cite{lsm}, and all parameters are
known exactly:
in units where $\hbar = k_B = 1$ we have
$g_c = J_1$, 
$N_0 = (1-(g/J_1)^2)^{1/8}$ and
$\epsilon_p = 2 ( J_1^2 + g^2 - 2 J_1 g \cos (pa))^{1/2}$,
where $a$ is the lattice spacing,
so
$mc^2=2(J_1-g)$.
However, the existence of these stable particles is {\em not}
a special feature of this integrable point.
Indeed, for $g \ll g_c$, a particle
has the simple
weak-coupling interpretation as the boundary between domains with opposite
orientations in $\langle \sigma^z \rangle$. Conversely, for $g \gg g_c$, the
ground states has all the spins oriented in the $+x$ direction, and the
particle is a $-x$ spin hopping from site to site. We expect that these
interpretations remain correct at the lowest energies, as we
approach $g_c$ from either side,
but not at $g=g_c$. 

In the above interpretations, the particles are evidently bosons.
They  have short-range repulsive interactions, which, by an elementary
calculation~\cite{korepbook} implies
that the two-particle $S$-matrix $S_{pp'}$ approaches $-1$ for
$pa, p'a \ll 1$. The integrable model has
$S_{pp'} = -1$ for all $p$, $p'$, 
and this is often used to obtain a free-fermion
description after a non-local gauge transformation: we shall not use this
transformation in our analytical computations.

Our new analytical results are in the two low $T$ regimes on either side of
$g_c$, with 
$T \ll |m c^2 |$. In these regimes~\cite{ssising}, the density, $\rho$, of
thermally excited particles $ \sim e^{- |mc^2|/T}$, is exponentially small,
their mean spacing is much larger than their thermal de Broglie wavelength,
$\sim (2mT)^{-1/2}$. These are the standard conditions for {\em
classical\/} behavior in which
$\rho = \int dp/(2\pi) e^{-\epsilon_p /T}$.

\begin{figure}
\epsfxsize=3in
\centerline{\epsffile{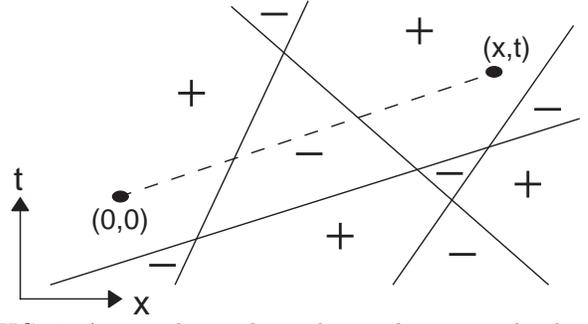}}
\caption{A typical semiclassical contribution to the double time path integral
for $C(x,t)$.  Full lines are thermally excited particles which propagate
forward and 
backward in time.  The $\pm$ signs are significant only for $g<g_c$
and denote the orientation of the order parameter. For $g>g_c$, the dashed line
is a particle propagating only forward in time from $(0,0)$ to $(x,t)$.}
\label{fig2}
\end{figure}

We now derive results in the ``renormalized classical'' (RC)
region $g<g_c$, $T \ll mc^2$.
Consider the correlation function
\begin{equation}
C(x_i ,t) = \mbox{Tr} \left ( e^{-H_I /T} e^{i H_I t} \sigma^z_i 
e^{-i H_I t} \sigma^z_0 \right)/Z ,
\label{keldysh}
\end{equation}
where $x_i = ia$, $Z = \mbox{Tr} e^{-H_I /T}$. 
This can be evaluated in terms
of the trajectories of the dilute gas of classical
particles noting that, since the particles
physically represent domain walls,
$\sigma^z$ changes sign every time a particle goes by.
Observe that
the classical trajectories remain straight lines
across collisions because the momenta before and after the
collision are the same in $d=1$.
This implies that the trajectories are
simply independently distributed straight lines, placed with a uniform
density $\rho$ along the $x$ axis, with an inverse slope
$ v_p \equiv d \epsilon_p /dp$, and with their
momenta chosen 
with the Boltzmann probability density $e^{-\epsilon_p /T}/\rho$ 
(Fig~\ref{fig2}).
This heuristic argument, based on the classical picture, can be justified
by taking the semi-classical (stationary phase) limit of the
the double-time (`Keldysh') path integral~\cite{keldysh}, in which 
each collision appears in
both the forward and backward paths (generated by the $e^{-iH_I t}$
and the $e^{i H_I t}$ in (\ref{keldysh}) respectively)
and therefore contributes 
the factor $|S_{pp'}|^2 = 1$.

Computing $C(x,t)$ is now an exercise in classical probabilities. 
The value of 
$\sigma^z (0,0) \sigma^z (x,t)$ 
is the square of 
the magnetization renormalized by quantum fluctuations ($N_0^2$),
times $(-1)$ if
the number of
trajectories intersecting the dashed line in Fig.~\ref{fig2} is odd.
Consider an Ising system of size $L \gg |x|$, and let it contain $N$ thermally
excited particles; then $\rho = N/L$.  Let the probability that any given
trajectory intersect the dashed line $= q$; then the probability only a given
set of $k$ lines will intersect is $=q^k (1-q)^{N-k}$.  Summing over all
possibilities, we have
\begin{eqnarray}
C(x,t) &=& N_0^2 
\sum_{k=0}^N (-1)^k q^k (1-q)^{N-k} N! /(k! (N-k)!) \nonumber \\
&=& N_0^2 (1-2q)^N \approx N_0^2 e^{-2 q N}.
\label{cxt}
\end{eqnarray}
The last step holds because, as we shall now compute, $q \ll 1$. First, for
$t=0$, as the density of trajectories along the $t=0$ axis
is uniform, $q = |x|/L$.  For $t \neq 0$, consider first all trajectories with
a fixed momentum $p$: they will intersect the dashed line if their intersection
with $t=0$ axis is between ordinates $0$ and $x -v_p t$, and so
$q = |x - v_p t|/L$.  Averaging over all $p$, and inserting in
(\ref{cxt}), we get one of our main results~\cite{maki}:
\begin{equation}
C(x,t) = N_0^2 R(x,t) \qquad\mbox{(RC region)} ,
\label{main_res_RC}
\end{equation}
\begin{equation}
\mbox{where}~~R(x, t)
= \exp\left( - \int \frac{dp}{\pi}e^{-\epsilon_p/T}
\left| x - v_p t \right| \right).
\label{rxt}
\end{equation}
Notice that 
$R(x,0) = e^{-|x|/\xi}$ and 
$R(0,t) = e^{-|t|/\tau}$, but the general behavior is
more complicated. The correlation length $\xi = 1/(2 \rho)$.
Remarkably, we find from (\ref{rxt}) that the correlation time, $\tau$, is
independent of the functional form of $\epsilon_p$ and depends only on the gap:
$\tau = (\pi/2T) e^{mc^2 /T}$.  In the $T \rightarrow 0$ scaling limit, in
which 
$\xi = (\pi/2mT)^{1/2} e^{mc^2/T}$,
$R(x,t)$
obeys the scaling form 
$R(x,t) = \phi \left( |x| /\xi, |t|/\tau \right)$,
where the scaling function $\phi$ is given by
\begin{equation}
\ln \phi(\bar{x}, \bar{t}) = - \bar{x}~\mbox{erf}
\left ( {\bar{x} \over\bar{t} \sqrt{\pi} }\right) -
 \bar{t}e^{-\bar{x}^2/(\pi \bar{t}^2)} .
\end{equation}
Similar classical scaling forms have been discussed earlier~\cite{ssising}, but
it was incorrectly conjectured that the scaling functions would be those of the
Glauber model~\cite{glauber}; Glauber dynamics does not conserve total
energy and momentum, and these conservation laws have played a crucial role in
the kinematic constraints on particle collisions.

\begin{figure}
\epsfxsize=2.75in
\centerline{\epsffile{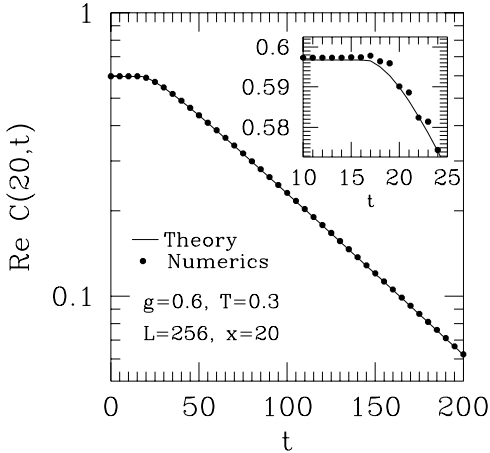}}
\caption{
The points show numerical data 
for the nearest neighbor model, in the RC
region, obtained for a lattice size $L=256$ with free
boundary conditions. This is compared with the theoretical
prediction in Eqs.~(\protect\ref{main_res_RC}) and
(\protect\ref{rxt}), in which the full lattice dispersion,
$\epsilon_p$ was used. In the numerics, the two sites were chosen to be as
close as possible to the center of the lattice. 
}
\label{fig3}
\end{figure}

We have also investigated the nearest neighbor model numerically by the mapping
to free fermions~\cite{lsm,mccoy}.
We generalized an earlier study~\cite{yr} of equal time properties to
dynamical quantities~\cite{snm} for the case
of free boundary conditions; details will be
presented elsewhere. We find that in the RC regime
the imaginary part of $C(x,t)$ is much smaller than its real part,
which suggests that the dynamics is indeed
classical as argued
above. In Fig.~\ref{fig3} we show data for $g=0.6, T=0.3$ and $x=20$ for a
lattice size $L=256$.
 We took $J_1=1$ so $g_c=1$ and $mc^2 = 2(1-g) = 0.8$.
For comparison we show the theoretical prediction from
Eqs.~(\ref{main_res_RC}) and (\ref{rxt}), in 
which we used the full lattice dispersion relation. The agreement is remarkably
good.  It is interesting to note that because there is a maximum velocity,
$v_{max}$, (on the lattice this is given by $v_{max}= 2 J_1 g$ for $g<g_c$ and
$v_{max}= 2 J_1$ for $g > g_c$, whereas in the continuum $v_{max}=c$),
Eq.~(\ref{rxt}) predicts that $C(x,t)$ should be independent of $t$ for $|t| <
|x| / v_{max}$, and the numerical results show this very clearly.  The inset,
with a much increased vertical scale, gives an idea of how small are the
deviations between the theory and numerics. We find that the agreement is also
excellent even at $r=0$, for $t>1$. 

Now we turn to the ``quantum disordered'' (QD) region 
$g> g_c$, $m<0$, $T \ll |mc^2|$. The
operator $\sigma^z$ flips spins between the $\pm x$ directions, and the large
$g$ picture, noted earlier, then suggests that $\sigma^z$ is the sum of a
creation and annihilation operator for the particles. As a result,
the $T=0$ the
spectral density, obtained from the Fourier transform of $C(x,t)$, has a
contribution $\sim \delta (\omega - \epsilon_p)$ associated with the stable
particle. Higher order corrections in $1/g$, or the form factor
expansion~\cite{formfac} on the continuum theory valid close to
$g_c$, show that
the next contribution to the spectral density is a continuum above the 3
particle threshold. Here we will focus exclusively on how the one particle
pole broadens as $T$
becomes non-zero. We define $K(x,t) \equiv C(x,t)$ at $T=0$, and dropping the 
multi-particle terms, we have
\begin{equation}
K(x,t) = \int \frac{dp}{2\pi} D(p) e^{i px - i \epsilon_p t}
\label{defk}
\end{equation}
where $D(p)$ is a form factor.  
For the general lattice model, $D(p)$ is not
known; neglecting the multi-particle terms, $D(p)$ is seen from (\ref{defk}) to
be the spatial Fourier transform of $C(x,0)$ {\em i.e.\/} the structure factor.
For the continuum theory, we have $D(p) = {\cal A} c/(2 \epsilon_p)$ where
${\cal A}$ is the dimensionless quasiparticle amplitude~\cite{formfac} (for an
underlying integrable lattice model ${\cal A} = 2 (|mc^2| / J_1)^{1/4}$),
whence 
$K(x,t) = {\cal A} K_0 (mc(x^2 - c^2 t^2)^{1/2})/(2\pi)$,
with $K_0$ the modified Bessel function.

Now we consider $T\neq 0$ in the semiclassical approximation. 
A typical set of paths 
contributing to the Keldysh path integral is still given by
Fig~\ref{fig2}, but its physical interpretation is now very different. The
dashed line now represents the trajectory of a particle created at $(0,0)$ and
annihilated at $(x,t)$, and $\pm$ signs in the domains should be ignored. In
the absence of any other particles this dashed line would contribute $K(x,t)$
to $C(x,t)$. The scattering off the background particles (the full lines in
Fig~\ref{fig2}) introduces factors of the $S$-matrix element $S_{pp'}$;
as the dashed line only propagates forward in time, the
$S$-matrix elements for collisions between the dashed and full lines (and {\em
only} these) are not neutralized by a complex conjugate partner. Using the low
momentum value $S_{pp'} = -1$, we see that the contribution 
to $C(x,t)$ equals $(-1)^{n_{\ell}} K(x,t)$ where $n_{\ell}$  is
the number of full lines intersecting the dashed line. The $(-1)^{n_{\ell}}$ is
precisely the term that appeared in the RC region, although for very
different reasons. We can carry out the averaging over all
trajectories as before, and obtain our 
main
result
\begin{equation}
C(x,t) = K(x,t) R(x,t) \qquad \mbox{(QD region)} ,
\label{main_res}
\end{equation}
where $K(x,t)$ is given by Eq.~(\ref{defk})
and $R(x, t)$ by Eq.~(\ref{rxt}). 
While (\ref{defk}) is valid for all $x, t$, relaxation
due to classical particles
makes sense only within the `light cone', and so strictly speaking, 
(\ref{main_res}) requires
$ct \gg x$. Outside the light cone, $K(x,t)$ decays exponentially to zero on
the short length scale $\sim 1/mc$ at which $R\approx 1$ and $T$-dependent
effects are not expected to be large:
so it is reasonable to use (\ref{main_res}) except, perhaps,
for $x,t$ extremely small.

The result (\ref{main_res}) 
clearly displays the separation in scales at which quantum and
thermal effects act. Quantum fluctuations determine the oscillatory, complex
function $K(x,t)$, which gives the
$T=0$ value of $C(x,t)$.
Exponential relaxation of spin correlations occurs at
longer scales $\sim \xi, \tau$, and is controlled by the classical motion of
particles and a purely real relaxation function $R(x,t)$.

\begin{figure}
\epsfxsize=2.75in
\centerline{\epsffile{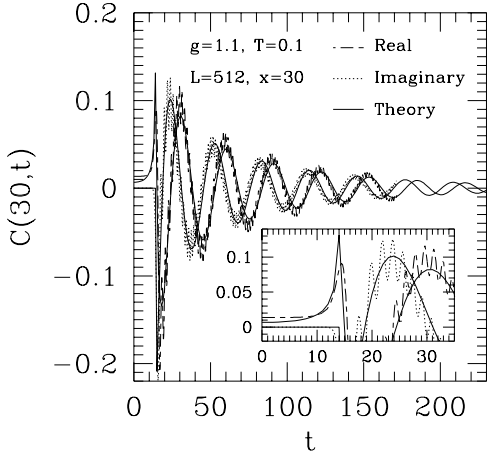}}
\caption{
The numerical data for the nearest neighbor model with $J_1=1$, in the
QD region,
obtained for a lattice size $L=512$ with free boundary conditions.
Also shown is the theoretical prediction from Eqs.~(\protect\ref{main_res}), 
(\protect\ref{rxt}) and 
(\protect\ref{defk}), with $D(p)$ determined in the continuum theory.
The numerical data has a ``ringing'' at high frequency, which we believe is due
to the upper energy cut-off in the dispersion relation, that is present in the
lattice model but not in the
continuum model used for the theory. The envelope of the numerical curve fits
the theoretical prediction
well (but not perfectly on the scale of resolution of the figure). 
The insert shows a part of the same data with an enlarged scale for clarity. 
}
\label{fig4}
\end{figure}

In Fig.~\ref{fig4} we
compare the predictions of Eq.~(\ref{main_res}) with numerical results on a
lattice of size $L=512$. 
The theoretical curve was determined
from the continuum expression for $K(x, t)$, but 
the full lattice form for $\epsilon_p$ was used in Eq.~(\ref{rxt}) to determine
$R(x, t)$.
With the parameters used, $|mc^2| =
0.2$, which is greater than $T$ $(=0.1)$ as required to be in the QD region.
The theory predicts that $\mbox{Im} C(x, t) = 0$ for $|t| < |x| / v_{max}$,
with a singularity 
in both real and imaginary
parts at $ |t| = |x| / v_{max}$. At longer time both 
parts oscillate and decay. Part of this decay comes from the
$1/\sqrt{t}$ decay of the Bessel functions, 
but for
$|t| > \tau$ this is dominated by the exponential decay from $R(x, t)$.
The theory agrees well with the numerics; some differences are
visible for small $x$, outside the light cone, but this is outside the
domain of validity of (\ref{main_res}). 

Precisely the same semiclassical arguments can also
be applied to the $d=1$ impenetrable gas of bosons of mass $m_B$, in a chemical
potential $\mu < 0$, with $T \ll |\mu|$.  This system has $S$-matrix elements
$S_{pp'} = -1$,
and its single 
particle propagator will be given by (\ref{main_res},\ref{rxt},\ref{defk}),
with $\epsilon_p = 
-\mu + p^2 /2 m_B$ and $D(p) = 1$. This propagator has also been computed
by the inverse scattering method~\cite{korepbook}, and 
the $T \ll |\mu|$ limit of their result agrees with
(\ref{main_res}). 

Turning to the ``quantum critical'' region $|mc^2 |\ll T \ll J_1,g$, the
dynamics is now given by finite $T$ correlators of the $g=g_c$ critical point.
An explicit expression for 
$C(x,t)$ was given earlier~\cite{perk,ssising}, and has
exponential decay at a single time scale $\sim 1/T$ in both its real and
imaginary parts.  There is no clear separation between the contributions of
thermal and quantum fluctuations since both are effective at this scale. Hence
there can be no effective classical model describing relaxation~\cite{CSY}.
For completeness, we also note the non-universal lattice high $T$ region $T \gg
J_1,g$; time-dependent correlators at $T=\infty$ were obtained
earlier~\cite{perk} and show a Gaussian decay in time.

Before concluding, we emphasize that although we compared the analytical
results with numerical data on an integrable model, the main result does not
depend on integrability. The simplification of the integrable model is that the
parameters in the theory, $m, c, {\cal A}$ and $N_0$, are known exactly,
whereas in general, they would be phenomenological parameters of a low energy
theory. The equality of the classical relaxation functions in the RC
and QD regions is surely related to the self-duality of the Ising critical
theory, and is a 
special feature of this model. Finally, it is interesting to speculate that
Eq.~(\ref{main_res}) may also 
be true for $d>1$ because of the separation of time scales between
the quantum and thermal fluctuations. 

We thank S. Majumdar, K. Damle, and T. Senthil for helpful discussions.
This research was supported by NSF Grants No DMR 96--23181  and DMR 94--11964.

\end{document}